\documentstyle[psfig,conf_iap]{article}

\def\ga{\mathrel{\mathchoice {\vcenter{\offinterlineskip\halign{\hfil
$\displaystyle##$\hfil\cr>\cr\sim\cr}}}
{\vcenter{\offinterlineskip\halign{\hfil$\textstyle##$\hfil\cr
>\cr\sim\cr}}}
{\vcenter{\offinterlineskip\halign{\hfil$\scriptstyle##$\hfil\cr
>\cr\sim\cr}}}
{\vcenter{\offinterlineskip\halign{\hfil$\scriptscriptstyle##$\hfil\cr
>\cr\sim\cr}}}}}
\def\la{\mathrel{\mathchoice {\vcenter{\offinterlineskip\halign{\hfil
$\displaystyle##$\hfil\cr<\cr\sim\cr}}}
{\vcenter{\offinterlineskip\halign{\hfil$\textstyle##$\hfil\cr
<\cr\sim\cr}}}
{\vcenter{\offinterlineskip\halign{\hfil$\scriptstyle##$\hfil\cr
<\cr\sim\cr}}}
{\vcenter{\offinterlineskip\halign{\hfil$\scriptscriptstyle##$\hfil\cr
<\cr\sim\cr}}}}}
\def\etal{{\it et al.} }
\def\eg{{\it e.g.} }

\begin{document}

\heading{%
%
EISily looking for distant galaxy clusters
%
} 
\par\medskip\noindent
\author{%
C. Lobo$^{1}$, D. Lazzati$^{1,2}$, A. Iovino$^{1}$, L. Guzzo$^{1}$, 
G. Chincarini$^{1,2}$
}
\address{%
Osservatorio Astronomico di Brera, via Brera 28, 20121 Milano, Italy
}
\address{%
Universit\'a degli Studi di Milano, v. Celoria 16, 20133, Milano, Italy
}

\begin{abstract}
We present a new algorithm to search for distant clusters of galaxies on 
catalogues deriving from imaging data, as those of the ESO Imaging Survey. 
We discuss the advantages of our technique relative to the ones 
developed before and present preliminary results of its application to the 
I--band data of the survey's patch A.
\end{abstract}

\section{Introduction}
The ESO Imaging Survey (EIS), presented in this conference \cite{Costa}, 
represents a unique opportunity of a large data set being 
made available to the community to produce all kinds of science. 
In particular, our working group at Brera is interested in the search 
for high redshift ($z \la 1.2$) galaxy clusters with the EIS-wide, as well 
as in the subsequent study of these systems and the evolution of 
their properties with redshift and environment (see \eg \cite{Stan} and 
references therein). Moreover, it is needless mentioning all cosmological 
questions that are raised simply by the existence and number density 
of clusters of galaxies up to very large distances and their use as
discriminant tests among different cosmological models (see \eg N. Bahcall's 
contribution in this conference).\\
The finding of a reliable set of cluster candidates in this or any other
catalogue produced by means of imaging data is, by itself, a preparation for 
the era of large telescopes such as the VLT, where we expect to perform the 
spectroscopic follow-up. This will, of course, confirm their real existence 
(and discard false chance alignments) and determine their distance/redshift. 
In fact, we shall have first an imaging campaign on 4m telescopes to 
produce photometric redshifts and to provide a first cut on individual cluster 
members. 
Large telescope time is precious, so we need to 
provide a list of robust candidates and this can only be accomplished 
if we have a highly performing cluster search method. So,
up to now, this has been our main objective: develop a cluster search
algorithm which gives a high success rate - that is, a high completeness 
level - without being overwhelmed by contamination in the form of spurious 
detections.\\
In their pioneering work on the Palomar Distant Cluster Survey, Postman 
and collaborators \cite{Post} wrote a matched filter algorithm 
to identify cluster candidates by using positional and photometric data 
simultaneously. We have tried to improve on their algorithm, especially in 
the sense of disposing of the {\it a priori} assumptions that are implicit 
in the Postman \etal technique.\\
Here we briefly describe our method, stressing the strong points and 
advantages of our algorithm, though leaving out the technical details, 
which can be found in a separate contribution by Lazzati \etal \cite{Lazz}. 
Finally, we shall present the results obtained by applying it to the 
preliminary EIS data of patch A that were released late march.

\section{Characteristics of our algorithm}
We use both spatial and luminosity information in
the form of a catalogue containing positions and magnitudes for all galaxies
detected in patch A. 
What we are looking for is a spatial concentration of galaxies which 
also presents a luminosity distribution typical of a cluster, say a Schechter, 
that hopefully sticks out from the background number counts - these ones 
following rather a power law.\\
The Postman algorithm relies on the choice of both a given cluster 
profile - modified Hubble - with a typical cluster size, and a typical 
M$^*$ (corresponding to the chosen parametrization of the Schechter). 
Moreover, the two parameters are fixed assuming a constraining relation 
between them, imposed by the tentative redshift that is thus assigned to 
each candidate.\\
Instead, in our new algorithm {\bf the spatial and luminosity part of 
the filter are run separately} on the catalogue and we make 
{\bf no assumption on the typical size or typical M$^*$ for clusters, as 
these parameters intervene in our algorithm as typical angular scale 
$\sigma$ and typical apparent magnitude m$^*$}.\\
For the spatial part we chose a {\bf gaussian filter} (a choice that 
brings along all the advantages and favorable properties of the Gauss function,
see \cite{Lazz}) and 
{\bf do not bind it to any fixed physical scale}. We rather use a set
of five widths - $\sigma$ - for the gaussian that cover a wide range of 
angular dimensions on the sky (from $\sim 0.35$ up to 
$\sim 1.42$ arcmin by steps of $\sqrt2$), instead of fixing an intrinsic 
value for the core radius (r$_C$) and a given cosmology.\\
Next to the spatial filtrage, we apply a maximum-likelihood technique 
to assess the signal given by the luminosity of the preliminary candidates
found in that first step. 
We use as functional form for the luminosity function a Schechter expressed 
in apparent magnitudes, thus avoiding the choice of an intrinsic M$^*$, and 
the choice of a given cosmology ($H_0$ and $q_0$) and morphological content 
(translated in the assumption of a given k-correction and, eventually, 
of an e-correction as well).\\
A very important point through all the process is that {\bf we estimate the 
background locally} for each candidate. 
This particular feature of our algorithm allows us to adapt well to and 
overcome the hazards of inhomogeneous data sets. Cluster candidates can 
thus be detected even in shallower regions of the catalogue.
In this way, we get the $\sigma$ and m$^*$ that independently maximize the 
signal with a given significance (respectively), combine these two 
and thus get a final probability for the detection of a cluster candidate.

Now, it is true that such a method does not allow us to produce, as a direct 
product and in a straightforward way, an estimate of the redshift for the 
cluster candidate. 
However, we believe that basing such an estimate on the rigid {\bf coupling} 
of r$_C$ with M$^*$ {\it via} redshift, as done in \cite{Post}, is a 
drawback~: not only does it bear unavoidable large uncertainties, but it 
also lowers the chance of detection of all candidates that do not flag a 
maximum 
likelihood at the very same redshift value simultaneously both for space and 
luminosity distributions. This is especially important when dealing with 
candidates close to the detection cutoff and knowing that the magnitude 
information is harder to deal with, providing less weight. Instead, our 
approach of combining the most probable m$^*$ with the most probable $\sigma$, 
produced in an independent way, allows us to preserve those candidates, 
thus reaching higher completeness values. Moreover, a redshift estimate is 
always feasible, in the ``standard'' way (and again with unavoidable 
uncertainties), through both these quantities.

\section{Preliminary results}
\begin{figure}[htbp]
\centerline{\vbox{
\psfig{figure=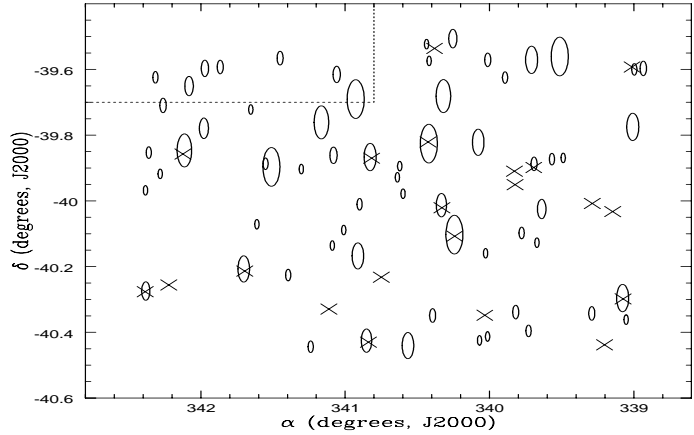,height=6.5cm,width=10cm}
}}
\caption[]{Map of the EIS patch A surveyed region. Rounds indicate 
$S/N \ge 3$ detections (the symbol size is proportional to the angular 
size of the filter for each case) obtained on both even and odd 
catalogues with our algorithm and with magnitude limit $I = 22.0$. Crosses 
note detections by \cite{Olsen} at the same S/N cutoff 
but for $I \le 23.0$.}
\label{fig:matches}
\end{figure}

Fig.~\ref{fig:matches} shows the results of our algorithm applied to 
both even and odd catalogues (see \cite{Costa} for these definitions) of the 
EIS patch A in the I band. This surveyed 
area totals $\sim3$ square degrees. The analysis has been limited 
to $I = 22.0$ so as to avoid data incompleteness, problems of discarding 
double objects at the borders of the individual frames and the steeply rising 
uncertainties of the star/galaxy separation at fainter magnitudes. Rounds 
indicate the candidates we found simultaneously in both catalogues with a 
cutoff S/N $\ga 3$. Our set totals $64$ cluster candidates. In order 
to assess our contamination rate we ran our algorithm on the original 
catalogues after shuffling galaxy positions (both in right ascension and 
declination) while maintaining their magnitudes. We got $\sim 1$ matched 
spurious detections in the whole patch area, which is actually a rough lower 
limit as it does not take into account the known angular two-point galaxy 
correlation function. 
In fig.~\ref{fig:matches} are also shown (crosses) the 21 detections obtained 
by \cite{Olsen}, again in both catalogues and with S/N $\ga 3$, but using the 
Postman \etal standard algorithm and a fainter limiting magnitude of 
$I = 23.0$. Moreover, in their candidate selection, they impose a cutoff 
in richness which most certainly accounts, at least partly, for the lower 
number of detections. Notice also that the spatial distribution of our 
candidates seems to be somewhat more homogeneous than in \cite{Olsen}, 
especially for $\alpha \ga 341^{\circ}$.2~
(even if \cite{Olsen} excluded from their analysis the upper left area, 
also marked in the figure).
This is very likely due to our different ways of 
estimating the background, our local assessment providing a clear advantage 
over their global fit throughout the whole patch. Also, it is clear that 
both these works have different objectives~: while we aim at obtaining a 
complete well-controlled sample (this being well patent in the philosophy of 
our algorithm), Olsen \etal are interested rather in 
providing reliable single candidates. And this would also contribute to us 
having a larger catalogue at the same cutoff, which also spans a wider range 
of richness classes. 
In common with \cite{Olsen}, we detect $11$ candidates ($68 \%$ of them having 
higher S/N in our catalogue). As for their remaining $10$~: (a) there are 
$6$ that we do detect but in one of the catalogues only, 
$2$ of which are 
flagged by Olsen \etal as ``problematic'' detections (see table notes in 
\cite{Olsen}); (b) and $4$ that we do not detect at all~: it is quite 
striking that, $2$ of these are labeled as {\it ``no obvious 
galaxy density visible''} by \cite{Olsen}.\\
Finally, we should add that, even if we adopt in this paper a cutoff of 
S/N $\ge 3$, a more reasonable value to be applied to our algorithm output 
would rather be $4$. This value lowers the total number of our matched 
detections to $26$, discarding the less rich and smaller angular sized 
systems. 
In this way, we still keep $10$ candidates in common with \cite{Olsen} 
(instead of the $11$ cited above) plus $3$ (instead of $6$) additional ones 
that we find in only one of the catalogues.

\acknowledgements{C. Lobo acknowledges financial support by the CNAA 
fellowship reference D.D. n.37 08/10/1997}

%

\vfill
\end{document}